\documentclass{aa}
\usepackage{graphicx}
\usepackage{txfonts}

\newcommand{\nsamp}{23,601 } 
\newcommand{\nnoise}{1180 } 
\newcommand{\ndata}{3203 } 
\newcommand{\nFP}{1122 } 
\newcommand{\Protmin}{1 }
\newcommand{\Protmax}{40 }
\newcommand{\Pcycmin}{0.5 }
\newcommand{\Pcycmax}{6 }
\newcommand{\Rvar}{R_{\rm var}}
\newcommand{\Prot}{P_{\rm rot}}
\newcommand{\Pcyc}{P_{\rm cyc}}
\newcommand{\Pcycvar}{P_{\rm cyc,\,var}}
\newcommand{\Pcycrot}{P_{\rm cyc,\,rot}}
\newcommand{\Pcycin}{P_{\rm cyc,\,in}}
\newcommand{\Pcycout}{P_{\rm cyc,\,out}}

\begin{document}
\title{Evidence for photometric activity cycles in \ndata Kepler stars}

\author{Timo Reinhold\inst{1,2}, Robert H. Cameron\inst{2}, Laurent Gizon\inst{2,1,3}}

\offprints{T. Reinhold, \email{reinhold@mps.mpg.de}}

\institute{
  Institut f\"ur Astrophysik, Georg-August-Universit\"at G\"ottingen, 
  37077 G\"ottingen, Germany
  \and 
  Max-Planck-Institut f\"ur Sonnensystemforschung,  
  Justus-von-Liebig-Weg 3, 37077 G\"ottingen, Germany
  \and 
  Center for Space Science, NYUAD Institute, New York University Abu Dhabi, PO Box 129188,
  Abu Dhabi, UAE
  }
  
\date{Received day month year / Accepted day month year}

\abstract
{
In recent years it has been claimed that the length of stellar activity cycles is 
determined by the stellar rotation rate. It is observed that the cycle period increases 
with rotation period along two distinct sequences, the so-called \textit{active} and 
\textit{inactive} sequences. In this picture the Sun occupies a solitary position in 
between the two sequences. Whether the Sun might undergo a transitional evolutionary 
stage is currently under debate.
}
{
Our goal is to measure cyclic variations of the stellar light curve amplitude and the 
rotation period using four years of Kepler data. Periodic changes of the light curve 
amplitude or the stellar rotation period are associated with an underlying activity cycle.
}
{
Using the McQuillan et al. 2014 sample we compute the rotation period and the variability 
amplitude for each individual Kepler quarter and search for periodic variations of both 
time series. To test for periodicity in each stellar time series we consider Lomb-Scargle 
periodograms and use a selection based on a False Alarm Probability (FAP).
}
{
We detect amplitude periodicities in \ndata stars between $\Pcycmin < \Pcyc < \Pcycmax$ 
years covering rotation periods between $\Protmin < \Prot < \Protmax$ days. Given our 
sample size of \nsamp stars and our selection criteria that the FAP is less than 5\%, 
this number is almost three times higher than that expected from pure noise. We do not 
detect periodicities in the rotation period beyond those expected from noise. 
Our measurements reveal that the cycle period shows a weak dependence on rotation rate, 
slightly increasing for longer rotation period. We further show that the shape of the 
variability deviates from a pure sine curve, consistent with observations of the solar 
cycle. The cycle shape does not show a statistically significant dependence on effective
temperature.
}
{
We detect activity cycles in more than 13\% of our final sample with a false alarm 
probability (calculated by randomly shuffling the measured 90-days variability 
measurements for each star) of 5\%. Our measurements do not support the existence of 
distinct sequences in the $\Prot-\Pcyc$ plane, although there is some evidence for the 
inactive sequence for rotation periods between 5--25 days. Unfortunately, the total 
observing time is too short to draw sound conclusions on activity cycles with similar 
length as the solar cycle.
}


\maketitle

\section{Introduction}
The origin of the 11-year solar activity cycle is one of the most important questions in 
solar and stellar physics. Over the course of one cycle the Sun undergoes phases of strong
and weak activity. During the active phase of the solar cycle, dark sunspots appear on the
solar surface. These spots have lifetimes of days to a few months \citep{Petrovay1997},
decaying into smaller magnetic concentrations which are bright (called faculae).
Individual faculae have short lives, as the magnetic flux associated with them is buffeted
by turbulent convective motions, however since magnetic flux is conserved, the faculae
reform and the lifetime of extended patches of faculae have much longer lifetimes. The
magnetic cycle of the Sun is thus accompanied by brightness variations. This variability
is weak during times of minimum activity, but the variability, being an evolving
combination of dark and bright features on the rotating surface of the Sun is high during
times of maximum activity. Although these activity phenomena act on different time scales
they can all be detected in the total solar irradiance (TSI) data (see e.g. \citealt{Domingo2009}).


On stars other than the Sun cyclic activity has also been observed through (long-term) 
brightness changes caused by epochs of increased occurrence of active regions on the 
surface or in the low stellar atmosphere. The magnetic field of the active regions
transports energy into the chromosphere, which leads to increased chromospheric emission,
most evident in the cores of the \ion{Ca}{ii} H\&K lines. The strength of this 
enhancement in the emission is usually described by the so-called \textit{Mount 
Wilson S-index} \citep{Vaughan1978} or the related quantity $R'_{\rm HK}$ \citep{Linsky1979}.
A long-term study of chromospheric activity of main-sequence stars in the solar 
neighborhood was initiated at the Mount-Wilson observatory in the late 1960s. 
\citet{Wilson1978} presented \ion{Ca}{ii} H\&K flux measurements with first evidence 
for cyclic variations. \citet{Vaughan1980} found that FGKM stars exhibit different levels 
of chromospheric activity with a deficiency of stars of intermediate activity. These authors 
explained the dearth of intermediate active stars in terms of an activity decline with age. 
In contrast, \citet{Noyes1984a} found a tight correlation between $R'_{\rm HK}$ 
activity and the Rossby number $Ro = \Prot/\tau_c$ (defined as rotation period $\Prot$ 
divided by the convective overturn time $\tau_c$), but could not detect evidence for the 
-- nowadays called -- \textit{Vaughan-Preston gap}. \citet{Noyes1984b} detected cyclic 
variations of the chromospheric activity levels in 13 main-sequence stars providing the 
relation $\Pcyc \propto \Prot^{1.25}$ between the cycle period and the rotation period. 
\citet{Baliunas1995} investigated stars of spectral type G0-K5 and found that 
these can roughly be separated by their mean \textit{S-index}: Fast rotators are assumed 
to be young objects with high \textit{S-indices}, whereas older stars rotate more slowly 
and show lower values of the \textit{S-index}, again providing evidence for the 
\textit{Vaughan-Preston gap}. \citet{Baliunas1996} tried to understand cyclic activity in 
terms of stellar dynamo theory, proving a relation between the ratio of the cycle and 
rotation period and the dynamo number $D$ according $\Pcyc/\Prot\propto D^{0.74}$. 
\citet{Brandenburg1998} and \citet{Saar1999} studied the relation between magnetic dynamo 
cycle period, rotation period, activity level and stellar age. These authors were the 
first claiming the existence of the \textit{active} (A) and \textit{inactive} (I) sequences, 
two almost parallel branches separated by a factor of six in $\Pcyc$ following the relation 
$\Pcyc/\Prot\propto\rm Ro^{0.5}$. Additionally, for fast rotators with $\Prot < 3$\,d 
these authors found evidence for a third branch with opposite slope 
$\Pcyc/\Prot \propto \rm Ro^{-0.4}$. \citet{BV2007} confirmed that chromospherically 
active stars follow the A- and I-sequences in the $\Prot-\Pcyc$ plane. Slow rotators 
exhibit shorter cycle periods populating the I-sequence, whereas stars rotating faster 
than the Sun exhibit long cycle periods located on the A-sequence. Some A-sequence stars
exhibit secondary cycle periods located on the I-branch. The Sun is located in between the 
two branches. In contrast, \citet{doNascimento2015} found that the Sun does not have a 
solitary position between the two branches but lies on a newly proposed solar-analog 
sequence.

All cycle period measurements above are based on cyclic \ion{Ca}{ii} H\&K emission, but also
long-term brightness variations indicate cyclic activity (see e.g. \citealt{Baliunas1985},
nicely linking solar activity to photometric and chromospheric activity on stars other than the Sun).
\citet{Olah2000} and \citet{Olah2002} investigated activity cycles from long-term V-band photometry. \citet{Messina2002} found evidence for cyclic activity in six solar-analogue stars also using
long-term V-band observations. In a subsequent paper, \citet{Messina2003} found that also
the mean rotation period changes over the course of the activity cycle. \citet{Lockwood2007}
provided parallel measurements of photometric and chromospheric variability, but unfortunately
did not determine any cycle periods. \citet{Olah2009} found a positive correlation between
rotational and cycle period for 20 active stars using time-frequency analysis. A similar approach
was applied to Kepler light curves of very fast rotators ($\Prot\leq 1$\,d) by \citet{Vida2014}.
These authors found hints for cyclic activity on timescales of 300--900~days analyzing cyclic
changes of the mean rotation period. Using CoRoT data \citet{FerreiraLopes2015} detected cyclic
variability in 16 FGK type stars. These authors found evidence for a possible third sequence,
parallel to the previously identified active and inactive sequences, but shifted by a factor
of $1/4$ towards shorter cycle periods. \citet{Lehtinen2016} found evidence for activity
cycles in 18 solar-type stars using decades-long photometric observations. The cycle periods
follow the sequences defined by \citet{Saar1999}, bridging chromospherically and photometrically
derived activity cycles. In recent years, activity cycles are also detected from asteroseismology
through periodic shifts in the mode frequencies \citep{Garcia2010, Salabert2016}.

The primary Kepler mission provides continuous, high-precision photometry of thousands of 
stars over four years of observation. Due to the presence or absence of active regions on 
the visible hemisphere (e.g. spots, plage) the stellar flux gradually changes. Such 
long-term changes are observed in the total solar irradiance data, showing significant 
changes over the course of the 11-year cycle. In addition, the preferred latitude of 
active region occurrence changes. Owing to differential rotation equatorial spots have a 
shorter rotation period than higher latitude spots. Our main goal is to detect cyclic 
changes of the light curve amplitude and the stellar rotation period, which we interpret 
as an indication of a photometric activity cycle. The largest sample with well determined 
rotation periods is provided by \citealt{McQuillan2014} (hereafter McQ14). This huge 
sample of $34,030$ active stars forms the basis of our activity cycle survey.

\section{Data \& Method}
\subsection{Motivation from the Sun}
\begin{figure}
  \resizebox{\hsize}{!}{\includegraphics{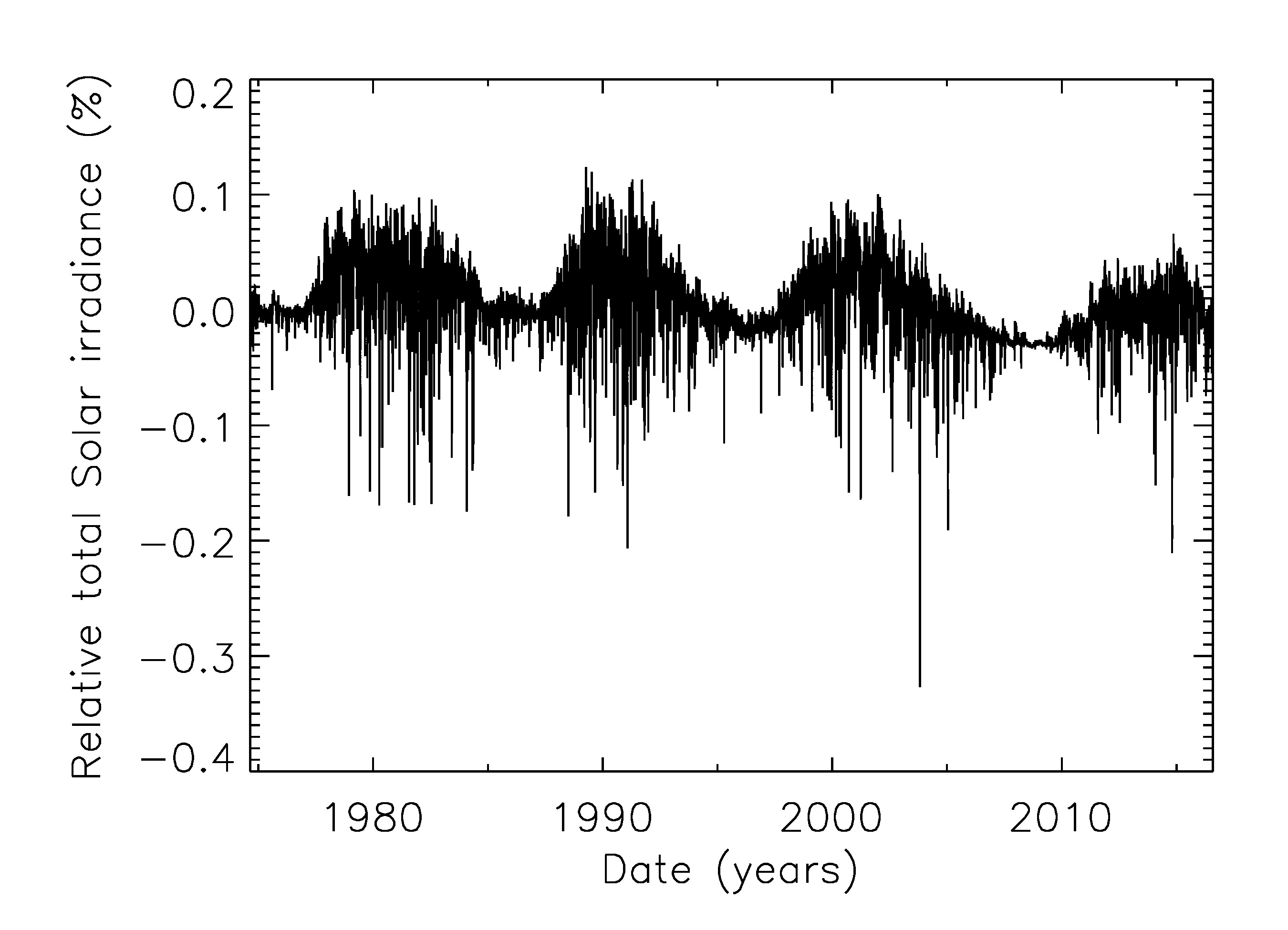}}
  \resizebox{\hsize}{!}{\includegraphics{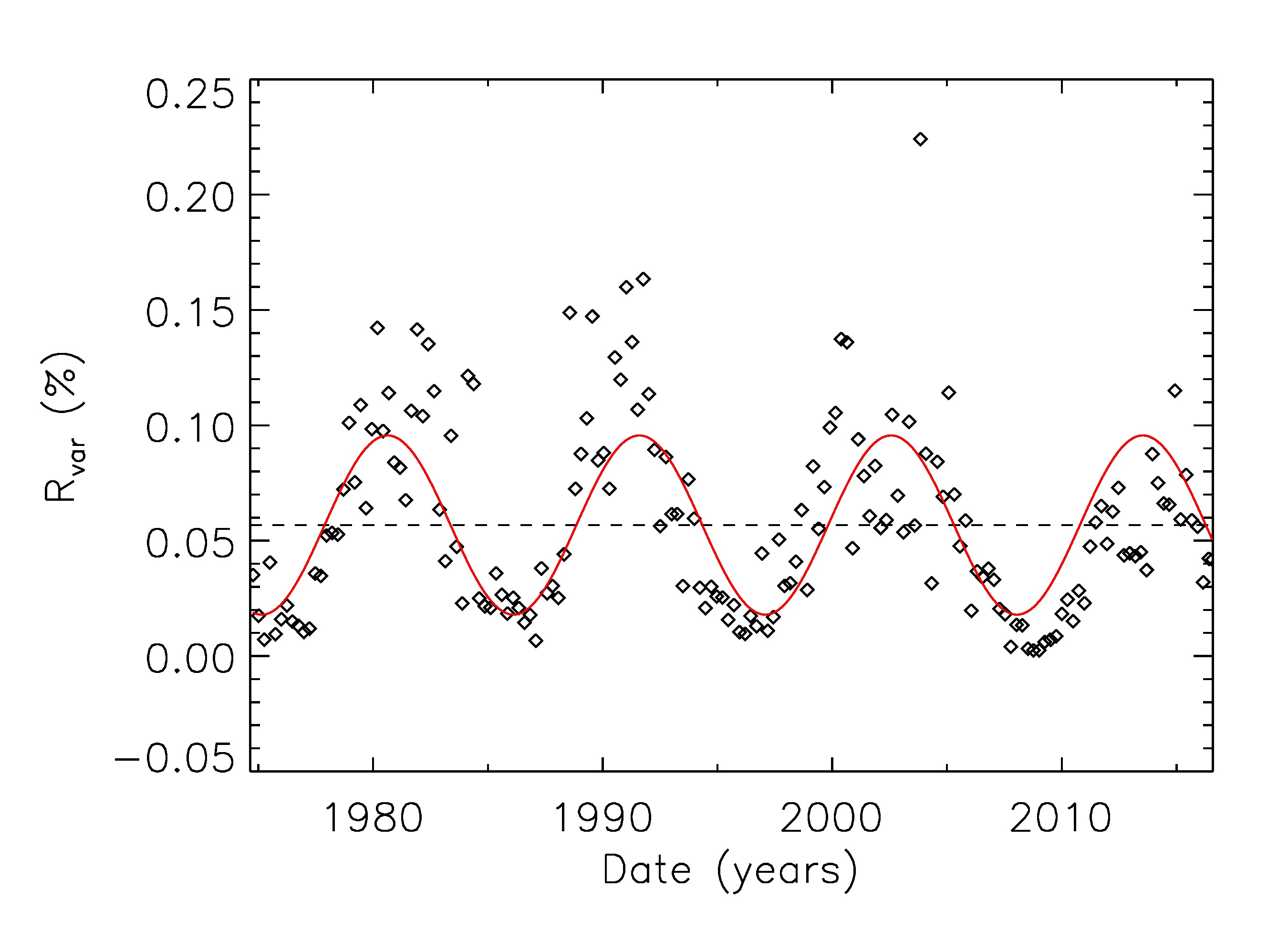}}
  \caption{Upper Panel: The measured Total Solar Irradiance (TSI), binned in 1-day 
  intervals, covering more than 40 years of observation. The approximately 11-year solar 
  cycle can be seen as the Sun's brightness changes from being almost constant during 
  activity minima to highly variable during activity maxima.
  Lower Panel: The variability range of the TSI data, $\Rvar$, as a function of time. 
  We define $\Rvar$ as the difference between the 5th and 95th percentiles of the 
  1-day binned intensity in each 90-day period in time. The red curve shows a sine fit 
  to the $\Rvar$ time series with a period $\Pcycvar = 10.98\pm0.26$~yr.}
  \label{TSI}
\end{figure}

As mentioned in the introduction, the brightness of the Sun (as measured by the total 
solar irradiance, TSI) varies due to the passage of dark sunspots and bright plage regions
across the solar disk. The top panel of Fig.~\ref{TSI} shows the TSI time series from 
\citet{Shapiro2016} produced with the SATIRE-S model \citep{Yeo2014}. It is readily seen 
that at times of activity maxima (1979, 1989, 2000, 2014) the TSI varies rapidly, while 
there are only small variations during times of activity minima (1976, 1986, 1996 and 2008).

To quantify the time-dependent variability in a manner which we can easily extend to Kepler
data we begin by rebinning the TSI data from a cadence of up to 1-minute to a 1-day cadence.
We then break the data into $N=73$ 90-days segments, and following \citet{Basri2011},
we define the variability range $\Rvar$ for that 90-day period as the difference between the
5th and 95th percentiles of the TSI measurements in the 90-day segment. This activity measure
is similar to the magnetic index $S_{\rm ph}$ defined by \citet{Mathur2014}. For each of the
$N=73$ segments we thus obtain a measurement of the variability $\Rvar(n), n=1,...,N$. Using
the midpoints of each 90-days time segment we create a time series $\Rvar(t)$, shown in the
lower panel of Fig.~\ref{TSI}. Computing the Lomb-Scargle periodogram of the time series
$\Rvar(t)$ reveals a peak at $\Pcycvar = 10.98\pm0.26$~yr, which is compatible with other
estimates of the solar cycle length. This shows that this method is capable of detecting the
correct cycle period from the variability of a stars light curve.

\subsection{Kepler Data}
Most Kepler data is released in segments of $\sim 90$ days (the so-called quarters) with 
exceptions for the quarters Q0 ($\sim 10$ days), Q1 ($\sim 33$ days), and Q17 
($\sim 32$ days). There is an ongoing debate how to remove instrumental effects from the 
time series. The data studied in this paper rely on the PDC-MAP pipeline using the 
following versions\footnote{The version number can be found in the primary headers of
the FITS files using the \textit{FILEVER} keyword.}: Version 2.1 for Q0-Q4 \& Q9-Q11; 
Version 3.0 for Q5-Q8 \& Q12-Q14; Version 5.0 for Q15-Q17. For each star we create a time 
sequence $t(n), n=0,...,17$, consisting of the midpoints of the time segments covered by 
the quarters $Qn$ when a star was observed.

\begin{figure}
  \resizebox{\hsize}{!}{\includegraphics{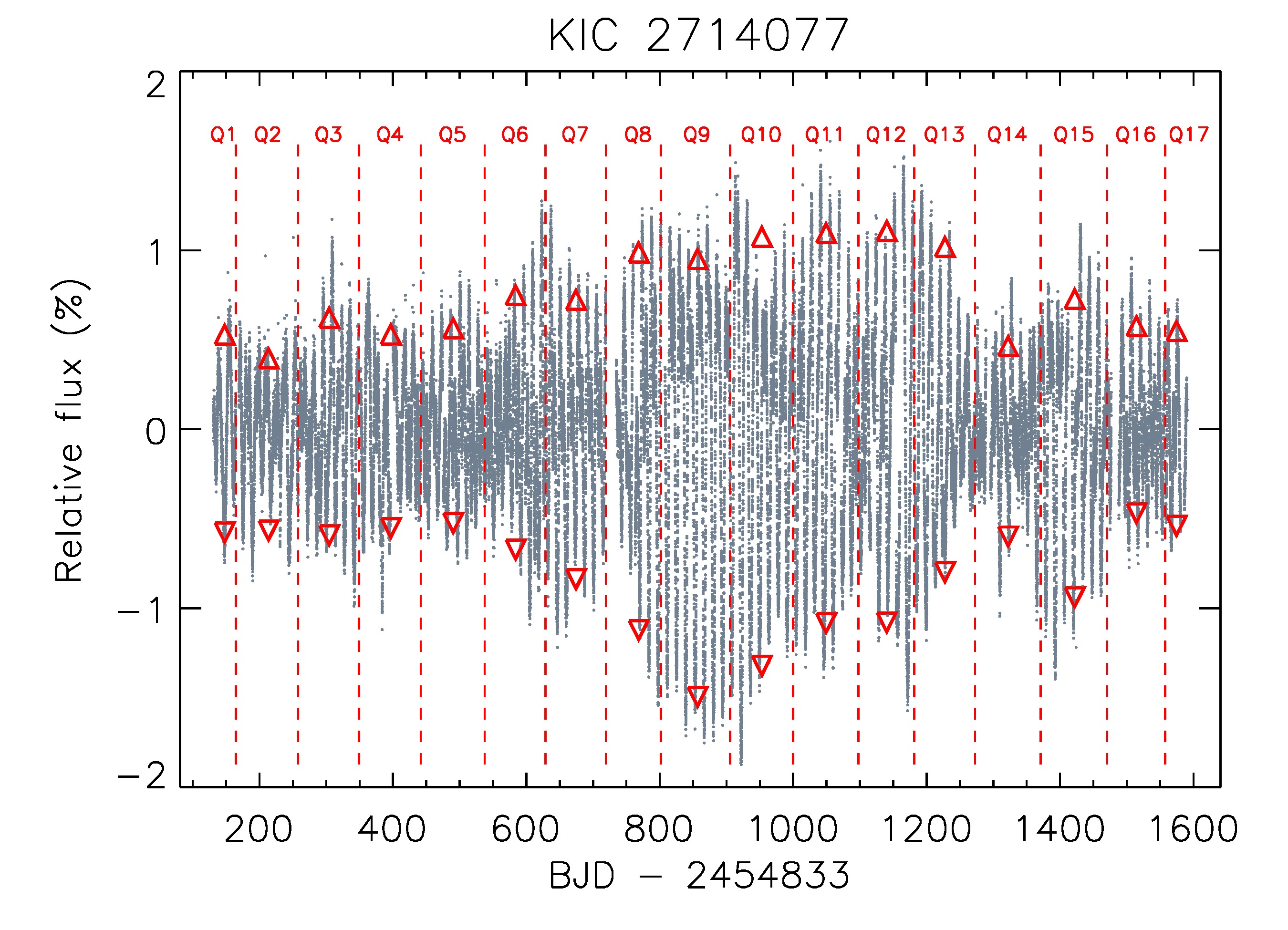}}
  \resizebox{\hsize}{!}{\includegraphics{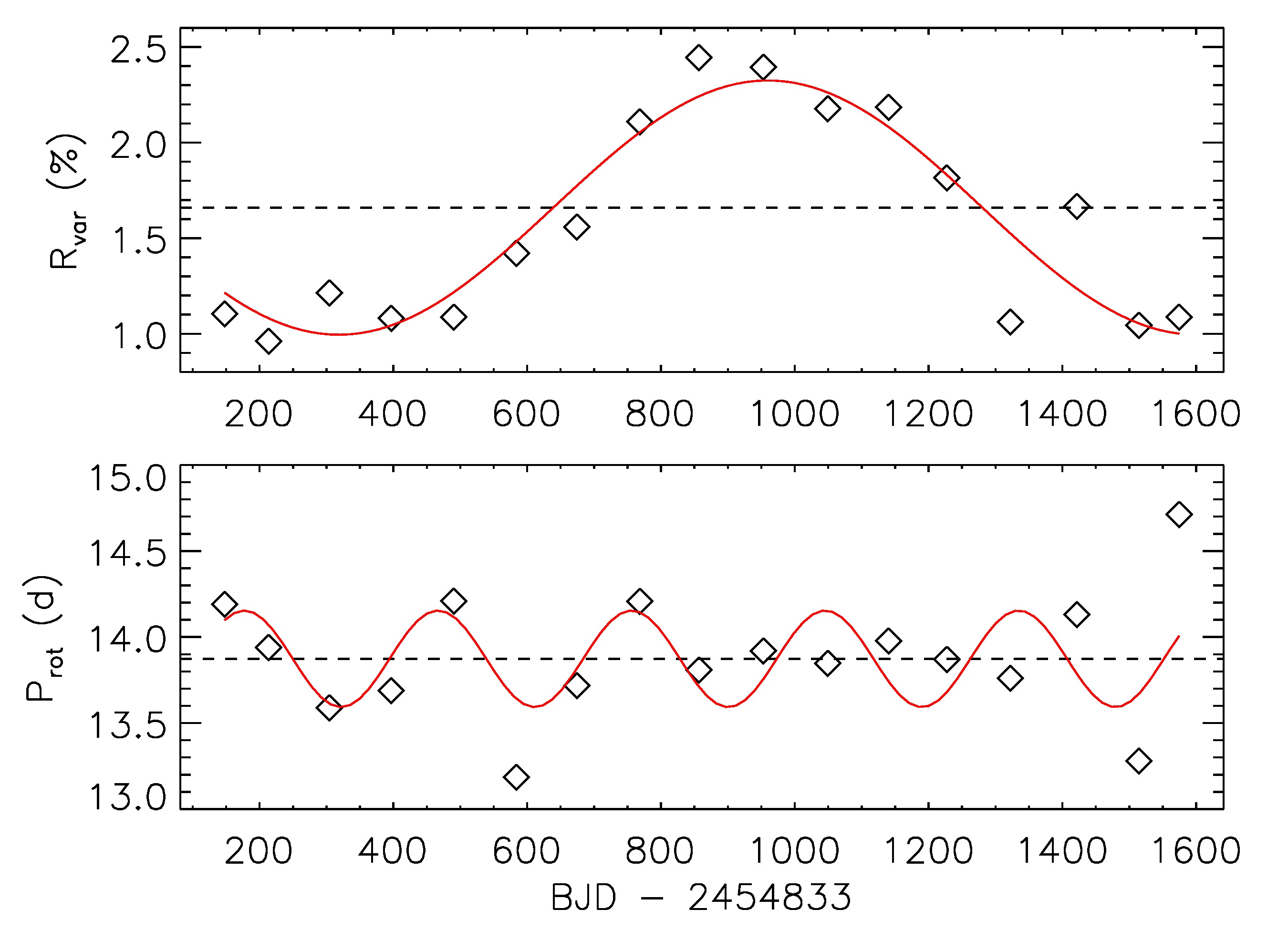}}
  \caption{Upper panel: Kepler light curve of the star KIC\,2714077. Vertical red lines 
  indicate the quarters Q1-Q17. Red triangles at the bottom and the top show the 5th and 
  95th percentile of the intensity, respectively. For each quarter we measure the intensity
  difference between the upper and the lower triangle which equals $\Rvar$.
  Middle panel: Time series of the variability range $\Rvar(t)$ of the same star, defined
  as the difference between the 5th and 95th percentiles in each quarter.
  Lower panel: Time series of the rotation periods $\Prot(t)$ of the same star. The red curve
  in the middle and lower panel shows the best sine fit to the $\Rvar(t)$ and $\Prot(t)$ time 
  series, respectively.}
  \label{lc}
\end{figure}

The upper panel of Fig.~\ref{lc} shows the light curve of the Kepler star KIC\,2714077. 
Each light curve is normalized and centered around zero by dividing the flux by its median
value and subtracting unity. This normalization is necessary because the Kepler
spacecraft rolls between consecutive quarters, so the stars fall on different CCDs accompanied
by offsets of the PDC-MAP flux between the quarters. Because the stellar variability we are
interested in is that due to the rotation of spots and faculae across the disk, we further fit
and subtract a second order polynomial from the data in each quarter. This will substantially
reduce the influence of possible longer term instrumental issues while only slightly affecting 
modulation comparable or faster than the rotation period of the star. We then use the 
$\Rvar$ defined in the same way as for the Sun as a measure of the variability. The red
triangles in the upper panel of Fig.~\ref{lc} indicate the 5th and 95th percentiles of the
relative flux in each quarter. The difference between the 5th and 95th percentiles equals
$\Rvar$, which is shown as time series $\Rvar(t)$ in the middle panel of Fig.~\ref{lc}.
The time series $\Rvar(t)$ shows strong periodicity with a period of $\Pcycvar = 3.51$~yr,
shown as red sine fit to the data.

In addition, to measuring $\Rvar$, for each star we also measure the stellar rotation 
period in each quarter using the Lomb-Scargle periodogram resulting in a time series 
$\Prot(t)$, which is shown for the same stars in the lower panel of Fig.~\ref{lc}. The 
quarterly measured rotation periods scatter around the mean rotation period $\Prot=13.88$~d.
In contrast to $\Rvar$ the time series $\Prot(t)$ does not show strong periodicity. The 
Lomb-Scargle periodogram finds a peak at a period less than one year, which is visualized by
the red sine fit to the data. As we will show in the following, strong periodicity of the 
variability range and random scatter of the rotation period time series is quite common for
many stars of our sample.

Owing to the limited quarter length we set upper limits on the detectable 
rotation periods: 10 days for Q0, 30 days for Q1 \& Q17, and 45 days for the other 
quarters. All period measurements deviating by more than 30\% from the median period of 
all quarters were discarded. Also the $\Rvar$ measurements of these quarters were 
discarded because instrumental artifacts might dominate the photometry. To achieve 
meaningful variability and rotation period time series we require measurements in at least
$N=12$ quarters. These conditions shrinks the initial McQ14 sample to $\nsamp$ stars. We
suspect that there might be more candidates with detectable activity cycle that were 
missed due to our constraints so far.

\subsection{Analysis of the time series}\label{analysis}
For each star we have 3 time series, $t(n)$, $\Rvar(n)$ and $\Prot(n)$ of length $N$, 
which is the number of quarters where a rotation period could be measured. The Lomb-Scargle
periodogram is a straightforward way to look for periodicities in unevenly sampled data.
We compute the Lomb-Scargle periodogram of the time series $\Rvar$ and $\Prot$ (using the
time series $t$) to test for periodicity. We call these periods associated with the
highest peaks in the Lomb-Scargle periodograms $\Pcycvar$ and $\Pcycrot$, respectively.

Because we are simply taking the highest peak in the Lomb-Scargle periodogram, $\Pcycvar$ 
and $\Pcycrot$ will be defined for every star. To define a subset of the stars where the
inferred periodicity is more likely to be real, we introduce a \textit{False Alarm 
Probability} (hereafter FAP) threshold. We calculate the FAP by considering 1000 random
permutations of $\Rvar$ and $\Prot$. For each permutation we have three time series 
$t$, $\Rvar^{\mathrm{Permutation}}$ and $\Prot^{\mathrm{Permutation}}$. We then create 
Lomb-Scargle periodograms for these new sequences and again concentrate on the highest 
peaks. We define the FAP for $\Rvar$ to be the number of cases where a higher peak 
appears in the Lomb-Scargle periodogram of $\Rvar^{\mathrm{Permutation}}$ than the one
which was found for $\Rvar$ divided by the number of permutations considered. 

\begin{figure*}
  \centering
  \includegraphics[width=17cm]{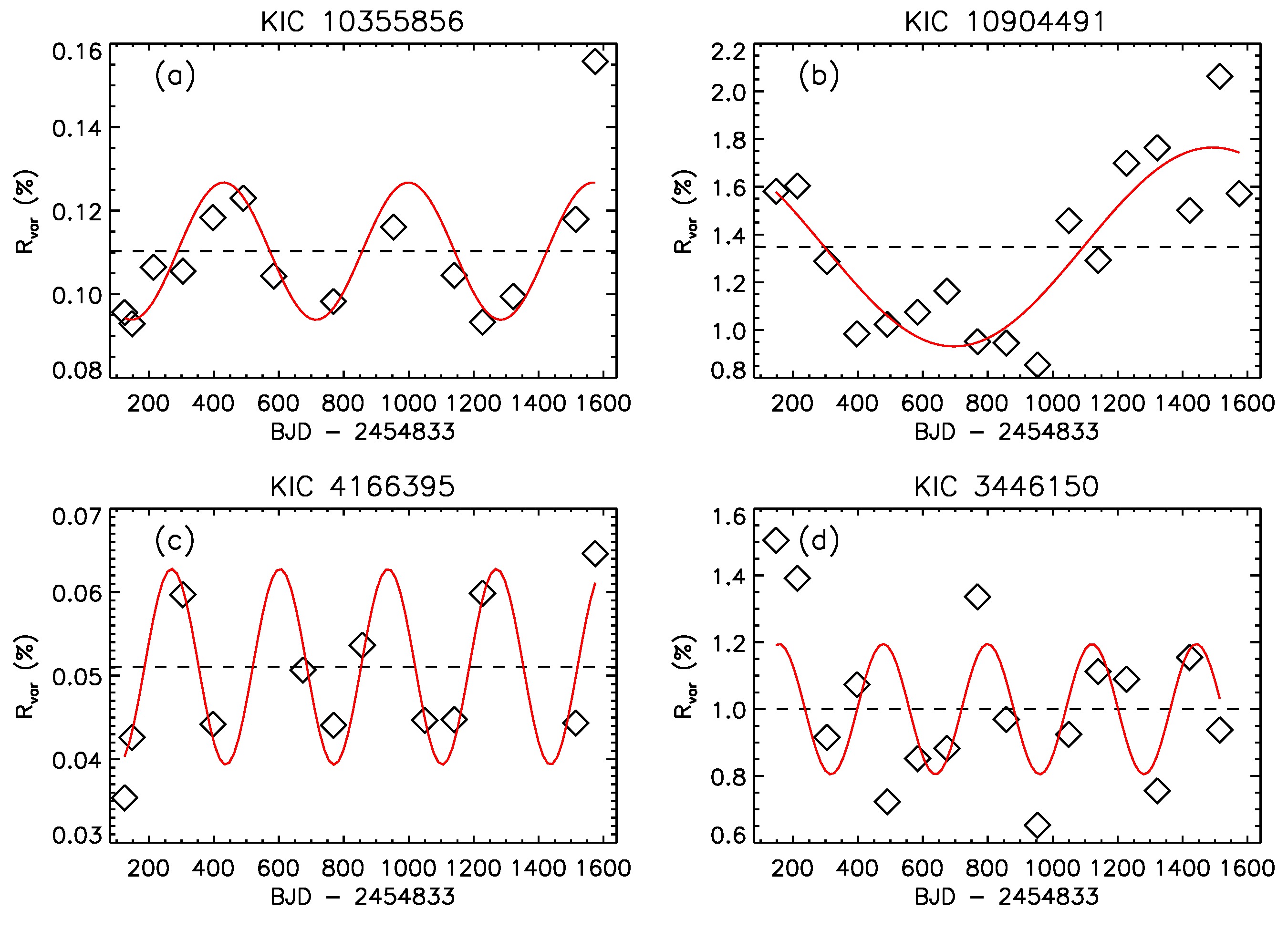} 
  \caption{Time series $\Rvar(t)$ of the quarterly measured variability ranges of four 
  Kepler stars. The time series in panel (a)-(c) pass our selection criteria that randomly
  permuting the time series $\Rvar$ of this star yields a higher peak in the Lomb-Scargle 
  periodogram in less than 5\% of all tested cases ($\rm FAP <5\%$), whereas panel (d) 
  shows a time series with ($\rm FAP >5\%$). The red curve shows the best sine fit to the
  data, and the dashed black line indicates the mean variability of the star. The $\Rvar(t)$
  time series shown in panel (a)-(d) exhibit false alarm probabilities $\rm FAP=1.0\%$, 
  $\rm FAP=0.5\%$, $\rm FAP=0.3\%$, and $\rm FAP=17.9\%$, respectively.}
  \label{examples}
\end{figure*}

We define the so-called \textit{periodic sample} by selecting those stars for which the
FAP was less then 5\%. The samples are defined independently for $\Pcycvar$ and $\Pcycrot$.
The time series $\Rvar(t)$ of three representative stars from the periodic sample
are shown in panel (a)-(c) in Fig.~\ref{examples}. For comparison panel (d) shows a star with 
$\rm FAP >5\%$. We discuss the issue of false positives in the results section.

\subsection{Method sensitivity}
\begin{figure}
  \resizebox{\hsize}{!}{\includegraphics{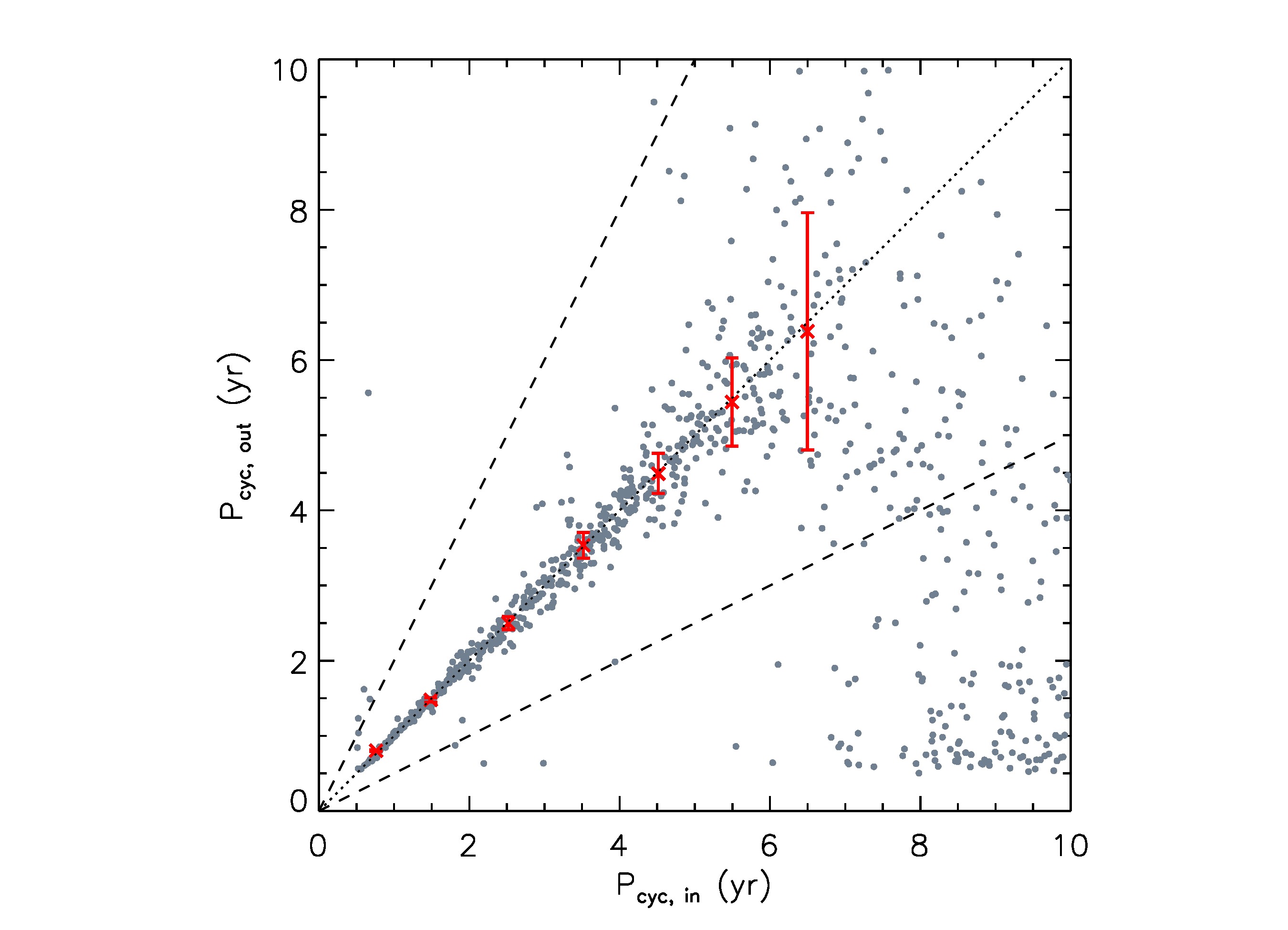}}
  \caption{Monte-Carlo test using a set of $1000$ simulated light curves showing the 
  input periods $\Pcycin$ and the returned periods $\Pcycout$. The black lines show 
  different $\Pcycout : \Pcycin$ period ratios: the dotted line shows the 1:1 period ratio,
  and the upper and lower dashed line show the 2:1 and 1:2 period ratios, respectively.
  The cycle period uncertainties (shown as red error bars) were estimated from
  ten different sets of $1000$ simulated light curves for each input period bin between
  0.5--10 years. The red crosses show the median value of $\Pcycout$ and the error bars
  show the median absolute difference between $\Pcycin$ and $\Pcycout$. We omit error bars
  for $\Pcycin>7$~yr because the uncertainties become huge.}
  \label{P_in_out}
\end{figure}

To test the sensitivity of the method to different cycle periods, we created a set of
$1000$ synthetic time series $\Rvar^{\rm syn}(t)$. We randomly selected $N$ data points,
$12 \leq N \leq 18$, from the time series $t(n)$. The time series $\Rvar^{\rm syn}(t)$ are
pure sine functions with input periods $\Pcycin$ uniformly distributed between
$0.5\,\rm{yr} < \Pcycin < 10\,\rm{yr}$. The amplitudes are uniformly distributed random numbers,
comparable to the sine fit amplitudes of the periodic sample. We add noise to the time
series $\Rvar^{\rm syn}(t)$, which we define as standard deviation of the difference between
$\Rvar(t)$ and the best sine fit from the periodic sample. The noise distribution is Gaussian.
Because the quantity $\Rvar$ describes a relative flux difference, the sine
amplitude and the noise are linearly correlated. For simplicity, the phase and the offset
of the sine curves are set to zero. We analyze the set of synthetic light curves using the
Lomb-Scargle periodogram, and compare the returned cycle periods $\Pcycout$ to the
input periods $\Pcycin$ in Fig.~\ref{P_in_out}. This Monte-Carlo test reveals that we can
trust cycle periods up to six years.

In the same way the uncertainties of the derived cycle periods have been estimated.
For different cycle period bins between 0.5--10 years we created a set of $1000$ synthetic
light curves for each bin, and applied the above analysis. In Fig.~\ref{P_in_out} the red
crosses show the median value of $\Pcycout$ and the error bars show the median absolute
difference between $\Pcycin$ and $\Pcycout$. We omit error bars for $\Pcycin>7$~yr because
the uncertainties become huge. The associated periods and uncertainties up to 
$\Pcycin<7$~yr are given in Table~\ref{P_io_table}.

\begin{table}
  \centering
  \begin{tabular}{cccc}
\hline\hline
$P_{\rm cyc,\,in}$ & $\langle P_{\rm cyc,\,out} \rangle$ & $\langle \Delta P_{\rm cyc,\,out} \rangle$ & $\langle \Delta P_{\rm cyc,\,out}/P_{\rm cyc,\,out} \rangle$ \\
(yr) & (yr) & (yr) & (\%) \\
\hline
$[0.5,1.0]$ & 0.80 & 0.01 & 1.2 \\
$[1.0,2.0]$ & 1.48 & 0.03 & 2.0 \\
$[2.0,3.0]$ & 2.50 & 0.09 & 3.6 \\
$[3.0,4.0]$ & 3.54 & 0.17 & 4.9 \\
$[4.0,5.0]$ & 4.49 & 0.27 & 6.1 \\
$[5.0,6.0]$ & 5.44 & 0.59 & 11.2 \\
$[6.0,7.0]$ & 6.38 & 1.58 & 26.2 \\
\hline
\end{tabular}

  \caption{Intervals of the input periods $\Pcycin$, median values of the output cycle 
  periods $\Pcycout$, and the associated absolute and relative uncertainties of $\Pcycout$
  as shown in Fig.~\ref{P_in_out}.}
  \label{P_io_table}
\end{table}

\section{Results}
\subsection{Cycle periods}
\begin{figure}
  \resizebox{\hsize}{!}{\includegraphics{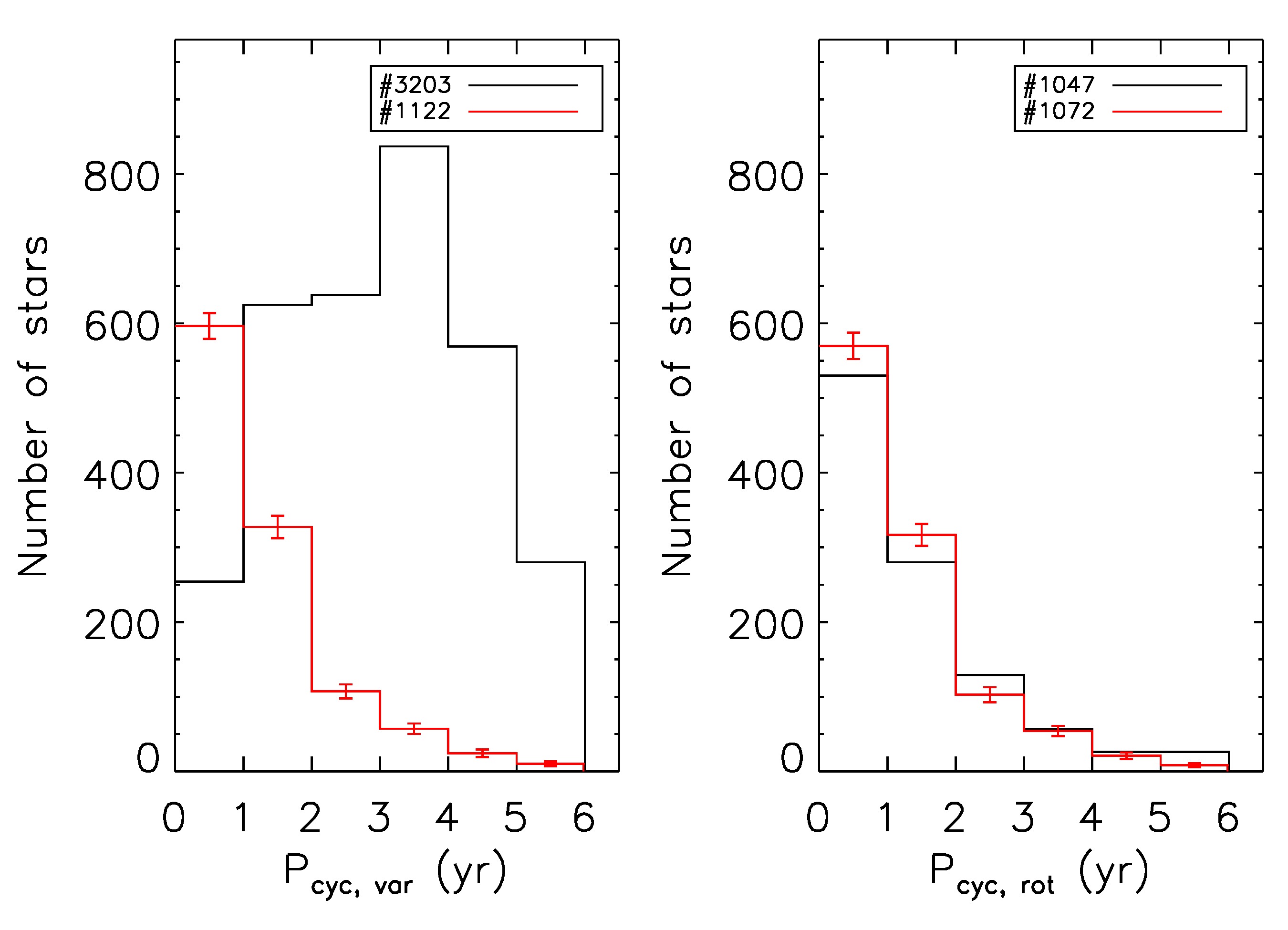}}
  \caption{Left panel: Distribution of cycle periods $\Pcycvar$ derived from all time 
  series $\Rvar$ which satisfied our selection criteria  $\rm FAP<5\%$. The black 
  distribution shows all cases where random permutations of time series $\Rvar$ yielded a
  higher peak in the Lomb-Scargle periodogram that the original time series in less than 
  5\% of all cases. The red curve shows the results from a Monte-Carlo experiment where 
  the analysis was performed on random permutations of each $\Rvar$ time series for the 
  whole sample. Error bars are based on the Monte-Carlo simulations and show one standard
  deviation.
  Right panel: Distribution of $\Pcycrot$, the cycle period determined from variations in
  the determined rotation periods $\Prot$, satisfying the selection criteria of $\rm FAP<5\%$.
  Again the black histogram shows the cycle period distribution of the 5\% strongest
  periodicity data, and the red curve shows the period distribution obtained from the 
  Monte-Carlo experiments.}
  \label{MCMC}
\end{figure}

The cycle period distribution of $\Pcycvar$ for all stars satisfying $\rm FAP<5\%$ 
is shown in black in the left panel of Fig.~\ref{MCMC}, and the distribution of $\Pcycrot$ 
for all stars satisfying $\rm FAP<5\%$ is shown in black in the right panel of
Fig.~\ref{MCMC}.
Assuming the time series $\Rvar(t)$ and $\Prot(t)$ do not contain any periodic signal, 
one would obtain a number of \nnoise stars after applying a limit of $\rm FAP<5\%$ to a 
total sample of \nsamp stars. For the cycle periods $\Pcycvar$ of the variability 
amplitude $\Rvar(t)$ we detect \ndata cycles. This number is significantly higher than the
expected number of false-positive detections.

To test whether the different number of detections might be an artifact of the method we
performed a Monte-Carlo simulation where we replaced the time series $\Rvar(n)$ of each 
star with a random permutation of that star's time series. We then applied exactly the
same analysis as was applied to the original time series (see Sect.~\ref{analysis}).

This randomization process is repeated 500 times for the whole sample. Comparing the peak
height of the initially randomized time series to the peak heights of the remaining ones 
yields the false alarm probability $\rm FAP_{MC}$ for each star. We repeated this process 
500 times, and found that on average \nFP stars satisfied our selection criteria. This is
consistent with the expectation from the condition $\rm FAP_{MC}<5\%$ applied to \nsamp stars. 
 
The averaged cycle period distribution from the Monte-Carlo experiment is shown in red in
Fig.~\ref{MCMC}, with error bars showing the standard deviation of different realizations.
It is obvious that the distributions of $\Pcycvar$ of the original and the randomized data
are very different. The distribution of the real data has a maximum around three years, 
whereas the randomized sample mostly contains short cycles between one and two years.

In the right panel the distribution of $\Pcycrot$ reveals a completely different picture. 
The distributions of the real data and the randomized sample are largely consistent, and 
the total number of detections is almost identical. This shows that we cannot distinguish
whether a cyclic change of the rotation period might indicate an activity cycle or if this
change occurs just by chance. We conclude that we cannot use the quantity $\Pcycrot$ to 
search for activity cycles. Thus, we concentrate on $\Pcycvar$ in the following.

\begin{figure*}
  \centering
  \includegraphics[width=17cm]{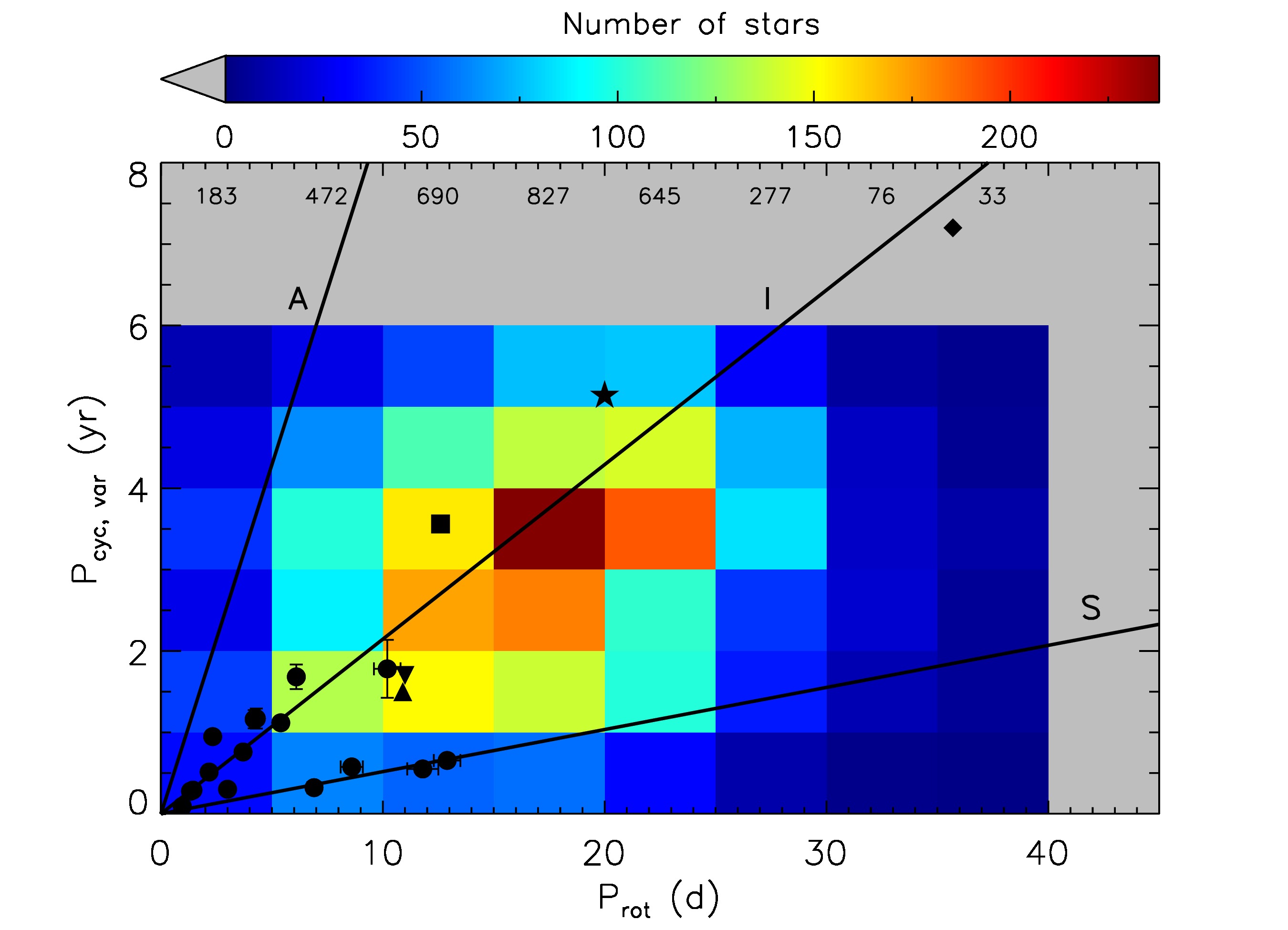}
  \caption{The color image shows the distribution of periodic stars with $\rm FAP<5\%$ in 
  the $\Prot-\Pcycvar$ plane, with gray indicating the limits of our detection 
  reliability. The different black symbols represent cycle period measurements from other 
  authors; circles: \citet{FerreiraLopes2015}, upside down triangle: \citet{Egeland2015},
  normal triangle: \citet{Salabert2016}, square: \citet{Moutou2016}, star: \citet{Flores2016}, 
  and diamond: \citet{BoroSaikia2016}. The solid black lines show the active (A), 
  inactive (I), and short-cycles (S) sequences as defined by \citet{FerreiraLopes2015}. 
  The numbers above the color bins denote the total number of stars in each column.}
  \label{density}
\end{figure*}

Fig.~\ref{density} shows the dependence of $\Pcycvar$ on the stellar rotation period 
$\Prot$ for all \ndata stars of the periodic sample. The number of stars in each bin on 
the abscissa is given in black above the color bins. 

We turn now to the issue of the cycle period's dependence on rotation period. Our 
observations only allow activity periods of less than six years to be detected. From
Fig.~\ref{density} it is obvious that most cycle periods have a length between 2--4 years
(compare Fig.~\ref{MCMC}), slightly clumping at rotation periods between 10--20 days. Up 
to rotation periods of 25 days the cycle period slightly increases with rotation period. 
For stars rotating even slower the cycle period behavior is hard to tell from the data. 
Using unit weights for all data points a linear fit yields 
$\Pcycvar = 0.04\,\Prot + 2.40$~yr, i.e. a weak dependence of the cycle on rotation period. 
This result is confirmed by the small Pearson correlation coefficient $r=0.20$ between the
two quantities.

Some previous authors have reported a correlation between the rotation period of a star 
and the period of its activity cycle, with periods preferentially lying on active and 
inactive branches (shown in Fig.~\ref{density}), and with some stars having detectable
periodicities of their activity cycles lying on both branches. Clearly our detection 
scheme can not find multiple activity periodicities, and we find that most of our 
detections are concentrated along the inactive branch, however the distributions are 
broad.

For comparison photometric cycle measurements of 16 CoRoT stars \citep{FerreiraLopes2015} 
are shown as black circles in Fig.~\ref{density}. The upside down triangle, normal triangle,
square, star, and diamond show measurements from \citealt{Egeland2015, Salabert2016,
Moutou2016, Flores2016, BoroSaikia2016}, respectively. \citet{Egeland2015} also detected a
longer activity cycle of $\sim 12$\,yr for the same star (HD\,30495).

\begin{figure}
  \resizebox{\hsize}{!}{\includegraphics{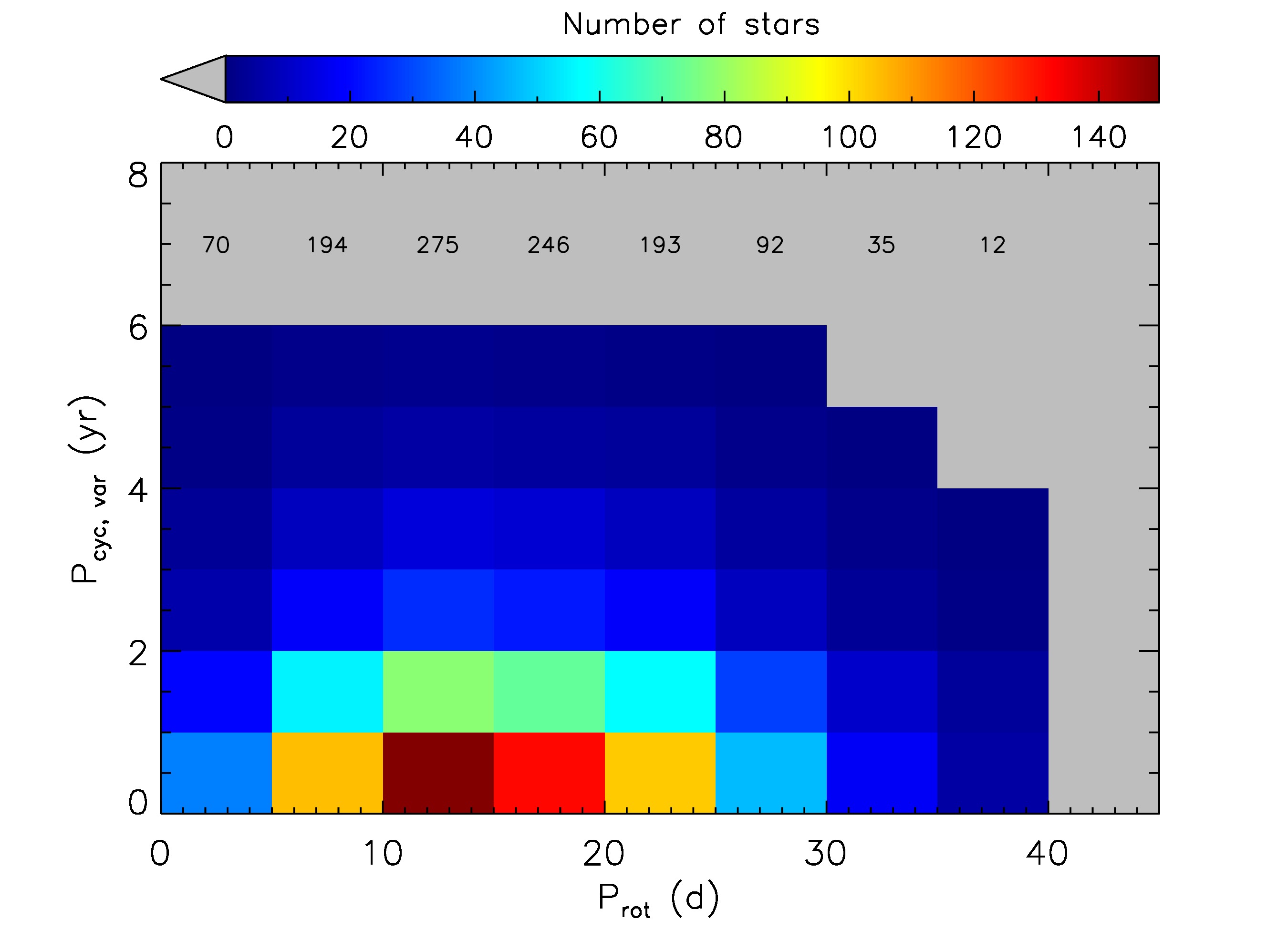}}
  \caption{Distribution of the cycle periods in the $\Prot-\Pcycvar$ plane expected from 
  noise (determined using Monte-Carlo experiments).}
  \label{density_MC}
\end{figure}

To check that Fig.~\ref{density} is not due to a hidden selection bias, we made the same 
figure beginning from the Monte-Carlo simulations discussed above. The result is shown in
Fig.~\ref{density_MC}. As expected from pure noise most cycle period detections 
lie in the range of 1--2 years because the periodogram interprets the noise as short
cycles. Furthermore, the sample shows a flat distribution, i.e., no dependence on rotation.
The associated Pearson correlation coefficient equals $r=0.01$, on average, showing that 
$\Pcycvar$ and $\Prot$ are uncorrelated.

\subsection{Cycle shapes}
\begin{figure}
  \resizebox{\hsize}{!}{\includegraphics{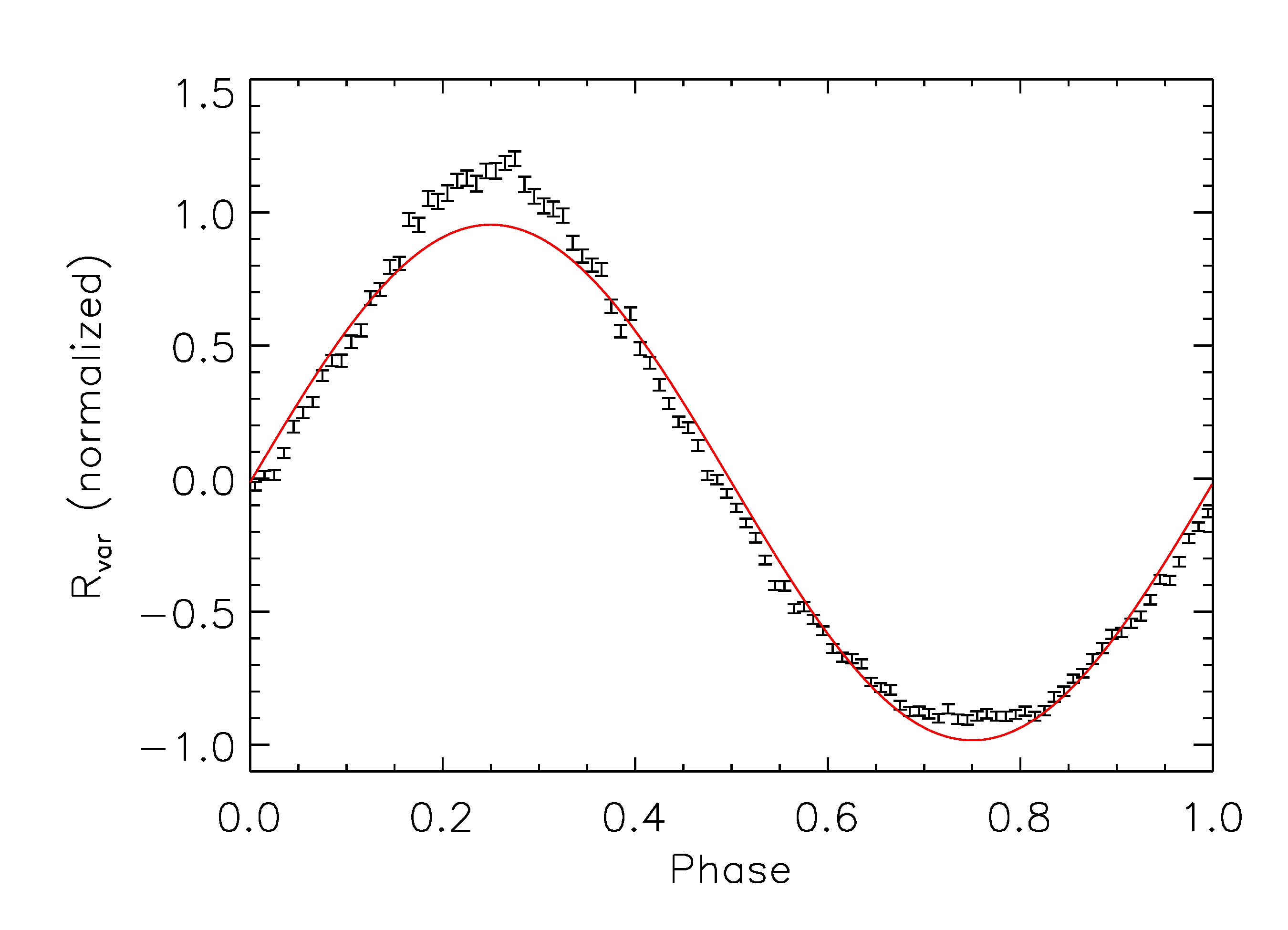}}
  \caption{Phase-folded $\Rvar(t)$ curves of all \ndata stars with strong periodicity. 
  Error bars show the standard error of the $\Rvar(t)$ distribution in each phase bin.}
  \label{phased}
\end{figure}

So far we focused on the cycle periods but have not considered the shape of the 
variability. Fig.~\ref{phased} shows a phase-folded curve of the variability range $\Rvar$
averaged over all stars of the periodic sample. All $\Rvar$ time series are phase-folded 
with the measured cycle period and shifted that they all lie in phase. The mean variability
level is subtracted, and the curves are normalized by their amplitudes. The average
over all curves is shown in black and the red curve shows a sine fit. The error bars show 
the standard error in each phase bin. It is clear that the shape of the variability shows
some deviation from a pure sine curve. At maximum activity the observed variability curve
is more spiky than a sine curve. At minimum activity the variability is rather flat.
Applying the same analysis to the $\Rvar$ time series of the TSI data (compare Fig.~\ref{TSI})
we also find a flat shape at minimum activity, but a large scatter at maximum activity. To
quantify the deviation of the data from the sine curve we compute the reduced chi-square 
and the associated p-value to be $\chi_{\rm red}^2=16.9$ and $p=0.002$, respectively,
indicating a high statistical significance for the deviation from a sinusoidal shape.

\begin{figure*}
  \centering
  \includegraphics[width=17cm]{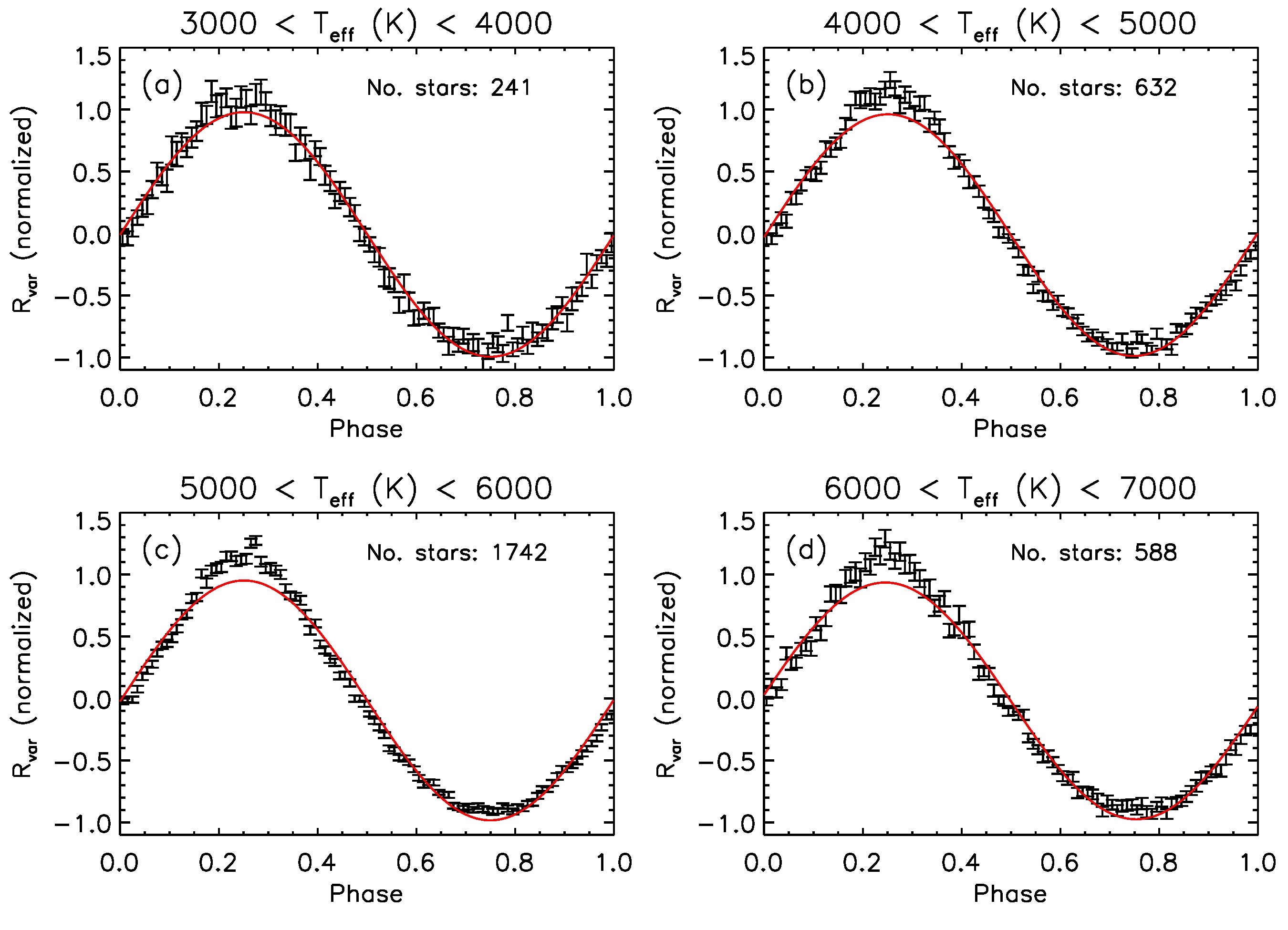}
  \caption{Phase-folded $\Rvar(t)$ curves of all periodic stars split up into different
  temperature bins as indicated on the top of each panel.}
  \label{phase_bins}
\end{figure*}

Fig.~\ref{phase_bins} shows the same curve as Fig.~\ref{phased} for different temperature
bins. In each bin the stars have roughly the same cycle period distribution.
The coolest stars are well represented by a sine curve. Neither the spiky top nor the flat
bottom are visible. The shape of the curve gradually changes towards hotter stars. The top
becomes more spiky and the bottom flattens out. To quantify the temperature dependence we
calculate the difference between each of the six possible pairs of the average curves (a)--(d)
in Fig.~\ref{phase_bins}. We do not find a statistically significant difference between the pairs
that would indicate a temperature dependence of the cycle period. The slightly different cycle
shape in panel (a) might be explained by the smaller number of stars in this temperature bin.

\section{Summary}
We detected periodicity of the variability amplitude $\Rvar$ in \ndata stars of the McQ14
sample, which we interpreted as indication for an underlying activity cycle. To account 
for random detections, the original sample was compared to a randomized sample. It was 
found that the number of activity cycle detections was three times higher than expected 
from pure noise. We did not detect cycles changes of the rotation period beyond those 
expected from noise. The $\Prot-\Pcyc$ plane revealed that the cycle period $\Pcycvar$ 
shows weak dependence on the rotation period, slightly increasing with rotation period.
In contrast to previous studies \citep{Noyes1984b, Baliunas1996, Saar1999, BV2007, 
FerreiraLopes2015, Lehtinen2016} we did not find evidence for a tight functional dependence
between the cycle and the rotation period. Our measurements show a large scatter around the
inactive branch. Previous studies (\citealt{Saar1999, BV2007}) have shown stars with multiple 
periodicities on the A- and I-branches. These studies were based on chromospheric activity
indices, which will have different sensitivities to spots and plage than the photospheric
variability detected using Kepler data. Hence it is possible that our finding of a strong
I-branch and little evidence of the A- and S-branches might indicate that the I-branch has
a stronger signature in the photosphere than the A- or S-branches. In good agreement with
the solar cycle the average shape of all activity cycles was found to deviate from a pure
sine curve. We could not detect a statistically significant temperature dependence
of the activity cycle shape.

\begin{acknowledgements}
The research leading to these results received funding from EU FP7 collaborative project
“Exploitation of space data for innovative helio- and asteroseismology” (SpaceInn) under 
grant agreement No. 312844. L.G. acknowledges support from the NYU Abu Dhabi Institute 
under grant G1502. This work was supported in part by the German space agency (Deutsches
Zentrum f\"ur Luft- und Raumfahrt) under PLATO grant 50OO1501. Part of the research 
leading to the presented results has received funding from the European Research Council 
under the European Community's Seventh Framework Programme (FP7/2007-2013) / ERC grant
agreement no 338251 (StellarAges).
\end{acknowledgements}

\bibliographystyle{aa}
\bibliography{biblothek}

\end{document}